\documentclass[conference]{IEEEtran}
\IEEEoverridecommandlockouts
\usepackage{cite}
\usepackage{amsmath,amssymb,amsfonts}
\usepackage{algorithmic}
\usepackage{graphicx}
\usepackage{textcomp}
\usepackage{xcolor}
\usepackage{color}
\usepackage{comment}
\usepackage{tabularx}

\usepackage{url}

\usepackage{caption}
\usepackage{subcaption}

\usepackage{wrapfig}

\usepackage{listings}
\lstnewenvironment{egdlisting}
{\minipage{\dimexpr\textwidth-2em}}
{\endminipage}

\usepackage{caption}
\usepackage{subcaption}

\usepackage{wrapfig}
\def\BibTeX{{\rm B\kern-.05em{\sc i\kern-.025em b}\kern-.08em
    T\kern-.1667em\lower.7ex\hbox{E}\kern-.125emX}}

\usepackage{comment}
\usepackage[normalem]{ulem}

\usepackage{color}

\begin{document}

\title{The SDSC Satellite Reverse Proxy Service for Launching Secure Jupyter Notebooks on High-Performance Computing Systems}


\author{\IEEEauthorblockN{
Mary P Thomas\IEEEauthorrefmark{1},
Martin Kandes\IEEEauthorrefmark{2}, 
James McDougall\IEEEauthorrefmark{3},
Dmitry Mishin\IEEEauthorrefmark{4}, \\
Scott Sakai  \IEEEauthorrefmark{5},
Subhashini Sivagnanam  \IEEEauthorrefmark{6} and
Mahidhar Tatineni  \IEEEauthorrefmark{7}
}
\IEEEauthorblockA{San Diego Supercomputer Center,
University of California San Diego, 
La Jolla, CA, USA\\
Email:\IEEEauthorrefmark{1}mpthomas@ucsd.edu,
\IEEEauthorrefmark{2}mkandes@sdsc.edu,
\IEEEauthorrefmark{3}jmcdouga@ucsd.edu,
\IEEEauthorrefmark{4}dmishin@sdsc.edu, \\
\IEEEauthorrefmark{5}ssakai@ucsd.edu,
\IEEEauthorrefmark{6}sivagnan@sdsc.edu,
\IEEEauthorrefmark{7}mahidhar@sdsc.edu}}

\maketitle

\begin{abstract}
Using Jupyter notebooks in an HPC environment exposes a system and its users to several security risks. The Satellite Proxy Service, developed at SDSC, addresses many of these security concerns by providing Jupyter Notebook servers with a token-authenticated HTTPS reverse proxy through which end users can access their notebooks securely with a single URL copied and pasted into their web browser.
\end{abstract}


\section{Introduction}
The Jupyter Project provides several tools and services including the Notebook Server (which launches Jupyter Notebooks, and JupyterLabs) and JupyterHub (which spawns, manages, and proxies multiple notebooks for multiple users) \cite{ProjectJupyterHome}. Because of their popularity, these tools and services are deployed on a large variety of High-Performance Computing Systems (HPC) resources \cite{Milligan-PEARC17, Zonca-PEARC18, Nicklas2018, Stubbs2020}. SDSC continues to support hosting Jupyter services on its HPC systems. However, as part of investigating the best methods for hosting JupyterHub and Jupyter Notebook services on SDSC HPC systems, we discovered several challenges and concerns in the area of security and resource usage issues. 

JupyterHub is a very popular, multi-user service which spawns, manages, and proxies multiple instances of Jupyter Notebook servers in a Web browser for multiple users. However, HPC system installation requires that the server be installed such that the service has access to the user home file systems. The security risks of hosting a JupyterHub on the login node of an HPC system, running as root, with thousands of user accounts presents a significant security risk and is not allowed by SDSC HPC system policies at this time. 

Deployment of notebooks on HPC systems has many challenges and requirements, especially when considering security. As part of our investigations, we found that users were launching notebooks from the command line, which by default are not secure (they are served over plaintext HTTP), and that users were not as protective of the URL, not realizing the vulnerability of putting the URL into the Web browser window. From the perspective of SDSC HPC system policies, the Jupyter Notebook is a backdoor into the user's account on the system; thus the URL to the notebook, in conjunction with the notebook's password, must be kept secret. In plaintext HTTP, it can be easily stolen by malicious parties eavesdropping on a user's network traffic. Using HTTPS with an untrusted certificate is not much better, as a malicious party can impersonate the legitimate Jupyter Notebook and acquire the user's notebook password and URL. In addition, as Jupyter Notebooks are essentially web pages, the host component of their URL requires some consideration, as to not create confusion (for people or browser security policy) with content officially served by the institution or with other users' notebooks. Launching notebooks with default configurations is by far the easiest for the user, the most popular, and the most insecure. In addition, users tend to launch these notebooks on the login nodes, which consumes resources needed by all the users who are logged onto those nodes.
\begin{figure*}[t]
\begin{subfigure}{0.48\textwidth}
\includegraphics[width=0.9\linewidth,height=4.5cm]{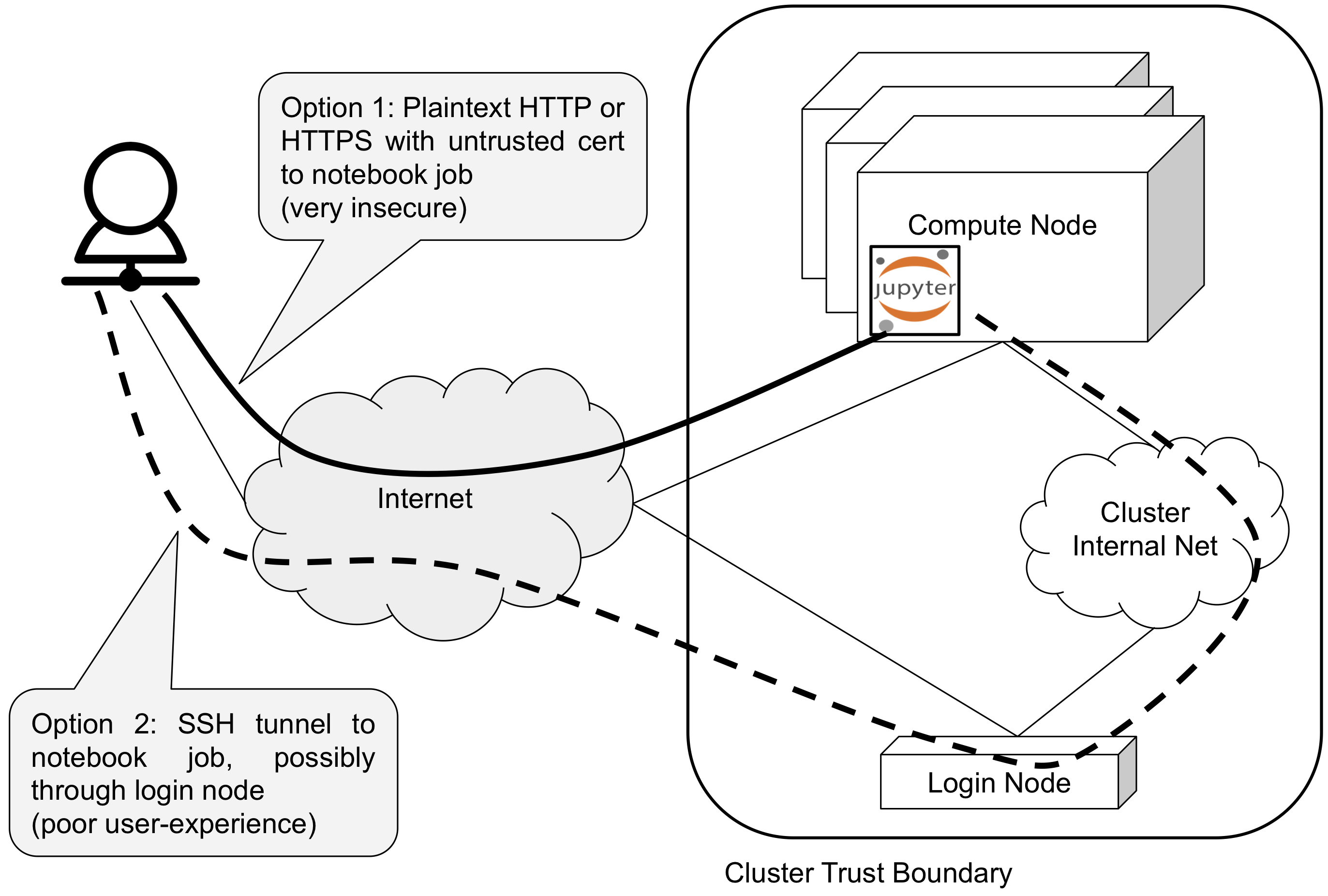} 
\caption{Jupyter Notebooks without the Satellite service.}
\label{fig:jup-ntbk-without-sat}
\end{subfigure}
\begin{subfigure}{0.48\textwidth}
\includegraphics[width=0.9\linewidth,height=4.5cm]{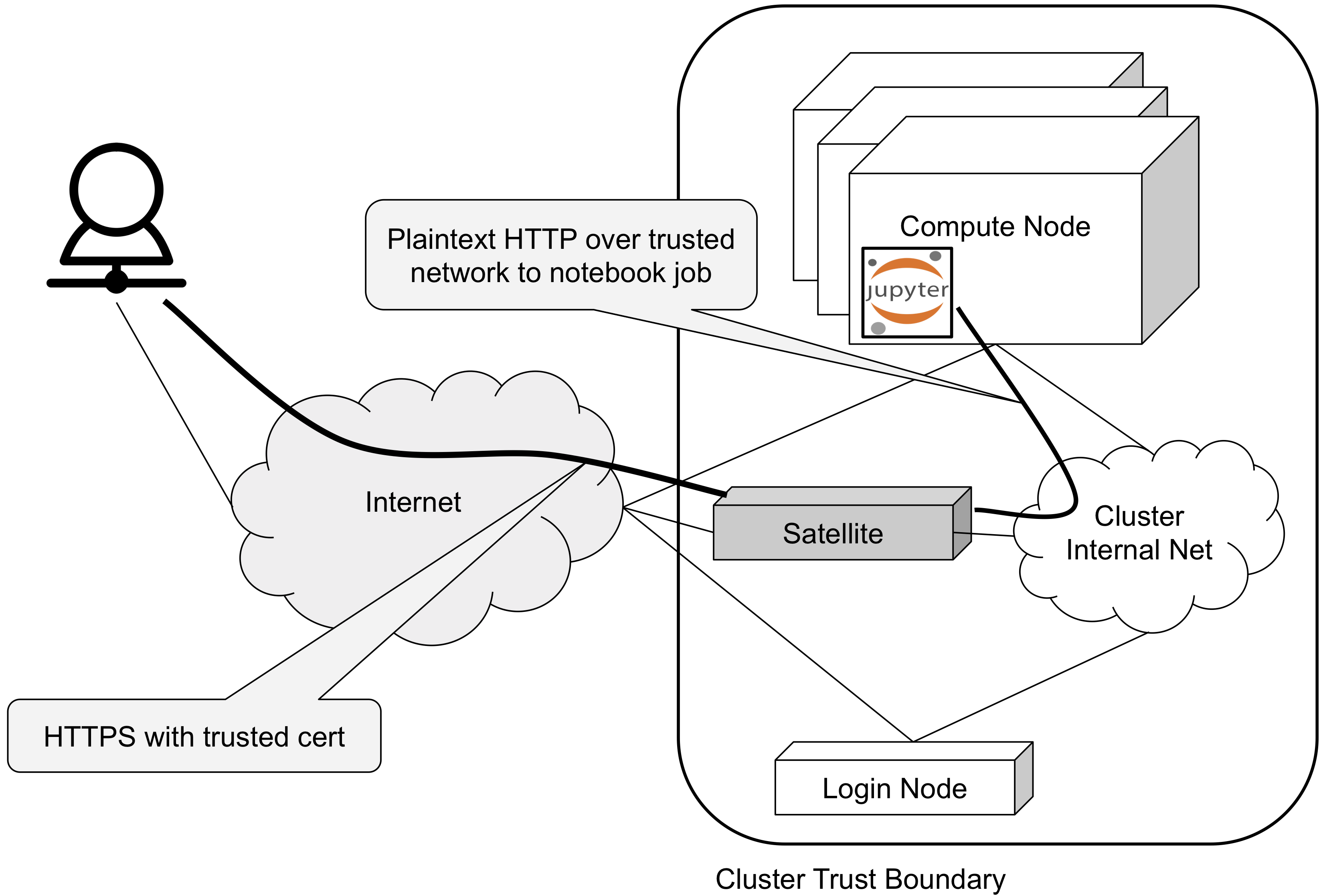}
\caption{Jupyter Notebooks with the Satellite service.}
\label{fig:jup-ntbk-with-sat}
\end{subfigure}
\caption{Jupyter Notebook deployment scenarios.}
\label{fig:jup-ntbk-sat}
\end{figure*}
Given the considerations described above, SDSC has a adopted a multi-tiered approach to running single-user notebooks more securely. We launched a campaign to educate uses about the risks and vulnerabilities of using Jupyter notebooks, encouraged users to run Jupyter Notebooks more securely over HTTPS, using self-signed certificates or using an SSH tunnel, and we began investigating more secure ways to deploy Jupyter Notebooks (along with or work on JupyterHub). In addition, SDSC HPC policies do not allow users to use the login nodes for computationally intensive processes, such as Jupyter notebooks. Computationally demanding jobs should be submitted and run through the batch queuing system. The two options for running secure notebooks mentioned above are represented in Figure \ref{fig:jup-ntbk-sat}(a), which shows the default deployment of Jupyter Notebook Services on an HPC system.

\textit{Option 1} uses plaintext HTTPS URLs, with untrusted SSL certificates to host the notebooks. A self-signed certificate can be generated by a user using \textit{openssl} and stored in a local file. However, the certificates must be signed by a certificate authority in order to be trusted by the user's browser. Not all users are willing to do the work required to set this configuration. Trusted certificates for each system on which the notebooks will be launched can be configured and made available for all users. However, this is not a long-term scalable solution when considering the number of nodes on an HPC system and the possibility of using the certificate for malicious purposes such as phishing.

\textit{Option 2} depicts using an encrypted SSH tunnel, which adds security by setting up all notebook communications to occur in the tunnel. This method is used successfully, but it is not stable as some firewalls and NAT gateways will forcefully terminate the SSH connection if it is idle fore more than a minute or two. Additionally, this method requires the user to determine the port of their notebook and configure their SSH client correctly. In instances where the node running the notebook does not have a network path to the user's browser, a second tunnel must be set up through an intermediate host, such as a login node. These factors make this method unattractive even though it sufficiently protects notebook-related traffic from being exposed on public or hostile networks.

However, these solutions are a challenge for users who are not computer scientists, and we have found that if you ask users to do something non-trivial and manually themselves to secure their notebooks, they will do the easy insecure thing they've always done. For standalone notebooks, not managed with a JupyterHub instance, we wanted to offer an incentive for users to serve their notebooks more securely by reducing the friction of the process. We needed a lightweight solution that would simplify the task of launching secure notebooks \text{by default}. The SDSC Satellite Proxy Service was developed at SDSC to address many of these security and usage concerns by providing Jupyter Notebook servers with a token-authenticated HTTPS reverse proxy, through which end users can access their notebooks securely with a single secure URL copied and pasted into their web browser.
%
%
%

\section{Architecture and Deployment}
The Satellite Proxy Server system is designed to simplify the process of launching a secure Jupyter Notebook by the client. The system consists of two main components: the Satellite Proxy Service and the Jupyter Spawner Client.

\subsection{The Satellite Proxy Service }
Figure \ref{fig:jup-ntbk-sat}(b) shows the architecture of the Satellite Proxy Service (Satellite). While the impetus for its creation is for proxying Jupyter Notebooks, Satellite was designed with the intent that it be used for any HTTP-based application with a similar mode of operation.

Satellite is comprised of an Apache httpd server, a handful of CGI scripts to manage mappings, and a cron job to generate httpd configurations for established mappings. The current version of Satellite requires access to two networks, a public network and a trusted internal network. Users interact with their application over the public network, and the trusted internal network is used for transiting the plaintext HTTP served by their application. It is worth noting that this internal network is already trusted for serving users' home directories over plaintext NFS, so this use of that network does not incur significant additional security risk.

Applications are served out of a unique, human-friendly URL, using a wildcard certificate signed by a CA/Browser Forum trusted CA, for a sub-domain unique to the Satellite deployment. To reduce the usefulness of Satellite for phishing attacks, we recommend that the subdomain be obviously labelled as user-generated content, and preferably unrelated to the institution's own domains. An example of a URL for one of SDSC's Satellite deployments is \textit{https://bullseye-compare-citation.comet-user-content.sdsc.edu}. The URL is recognized as soon as it is created, however there may be a long delay until the user's job starts and their application completes the process to establish a mapping. The user's job may also not run, or fail to run, creating a situation indistinguishable from a long wait time. To help address this case, an optional external component may submit (via HTTP POST) JobID-status information to Satellite, which will then attempt to provide more useful information about how the user's job is progressing through the batch queue when visiting their unique, but un-mapped URL.

\begin{figure}[t]
  \centering
  \includegraphics[width=.8\linewidth]{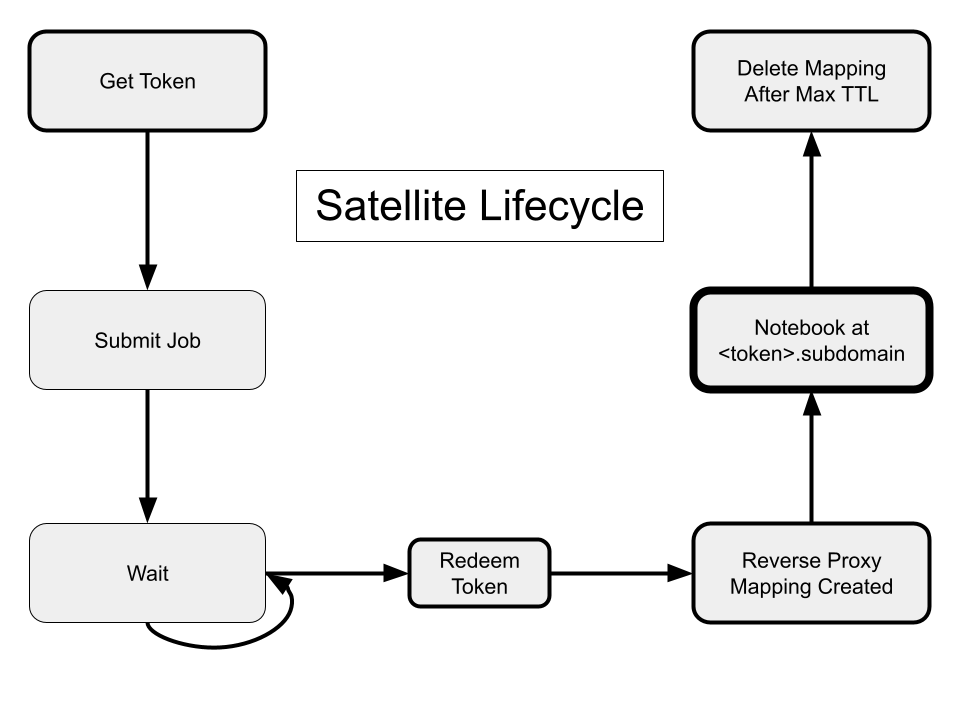}
  \caption{Sequence diagram for the Satellite Jupyter Notebook Proxy Service.}
 \label{fig:satellite-lifecycle}
\end{figure}

The operation of the Satellite service is shown in the Satellite lifecyle diagram shown in Figure \ref{fig:satellite-lifecycle}. First, from the login node, the user (or a script run by the user) obtains a \textit{token} by making an HTTP request to the \textit{getlink.cgi} script. The unique URL for the mapping will be at \textit{https://<token>.<satellite subdomain>}, and can be loaded in the user's web browser immediately. The token must be temporarily stored in a place where the user's batch job can retrieve it once it is run (such as a temporary file in their home directory or environment variable), and the user may now submit their batch job. When the user's job starts on a compute node, the job will involve scripting to determine the port number of the application, retrieve the token, and make a request to the \textit{redeemtoken.cgi} script, passing the token and port number as arguments. Satellite will create a mapping using the IP of the client and supplied port number. A cron job will put the mapping into effect, at which point the unique URL will return proxied content instead of a placeholder. The mapping will be removed after a configured amount of time, not to exceed the wall-clock limit of batch jobs for the system. To remove the mapping earlier, the \textit{destroytoken.cgi} script may be used from the same host the mapping was created on, with similar syntax to \textit{redeemtoken.cgi}.

To further reduce the possibility of misuse, Satellite requires token management requests to originate from the cluster's internal (trusted) network, and will only establish mappings to IPs in the cluster's internal network. Further, Satellite will not proxy to privileged ports (<1024). The use of unique subdomains ensures that the user's browser's security policy prevents trivially leaking cookies or HTTP authentication from one proxied application to another.

\subsection{Jupyter Spawner Client}
The Jupyter Spawner Client wraps up the above process of interacting with Satellite and preferably hides it from the user: its primary output being a URL and possibly additional password needed to access the application.


This process may be further streamlined by having a centrally-managed copy of \textit{start-jupyter} as well as incorporating its use in pre-built container images.

The goal of the \textit{start-jupyter} client is to spawn a Jupyter Notebook on various HPC Systems by using the Satellite service. The client software system includes: the \textit{start-jupyter} script; batch script templates  different queuing systems; a configuration file containing information about the HPC systems that are supported; library routines used for interacting with Satellite. Clients can use these pre-built batch scripts or customize them as needed. 
On launch, the \textit{start-jupyter} script will read the configuration file, load in the variables by sensing what HPC system the user is running on, check that the conda and jupyter environments are loaded, contacts the Satellite endpoint for tokens and to register the JobID, builds and submits the batchscript, prints out the URL to the user.

To run \textit{start-jupyter} the user needs to \textit{cd} into the repository. The \textit{start-jupyter} command has default values configured, but the user can customize several key variables including the project number, the batch script name, the jupyter service type (notebook, jupyterlab), the partition, number of GPUs, and other inputs.
\begin{figure*}[t]
  \centering
  \includegraphics[width=.6\linewidth]{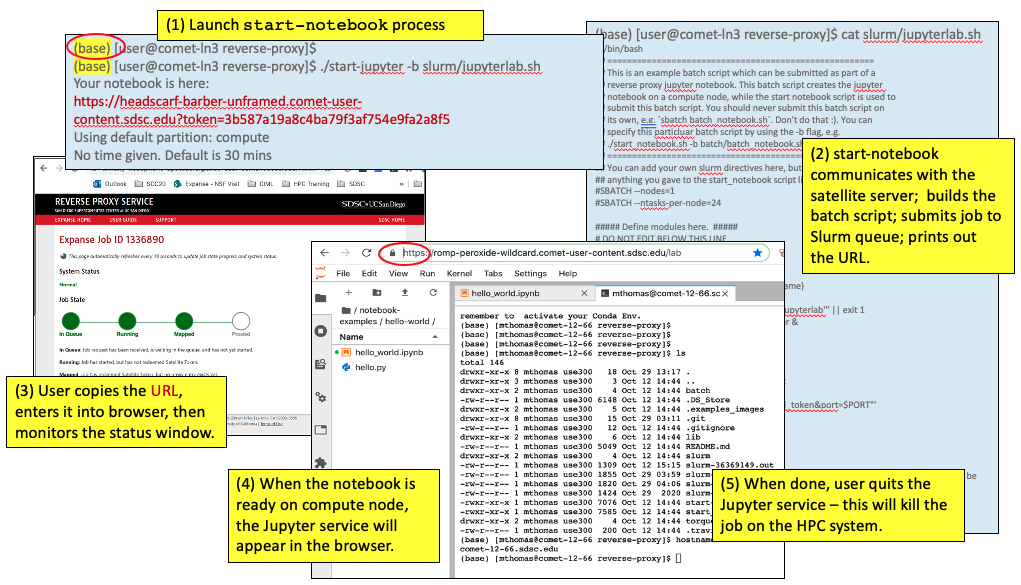}
  \caption{Workflow for running notebooks using the Satellite service}
  \label{fig:workflow}
\end{figure*}
\section{Using Satellite}

Satellite client scripts are intended to be easy on the user. After cloning the client repository, the user can run the \textit{start-jupyter} as is, or he may wish to modify a batchscript, or call the script with customized settings for the different variables. 

Figure \ref{fig:workflow} depicts how the \textit{start-jupyter} client is used to access the Satellite system to spawn a notebook on an Expanse CPU. The user initiates the process by running \textit{start-jupyter} which does the following: (1) contacts the Satellite server for a token, prints out the URL; (2) builds the batch script based on default setting and arguments passed via the command line, then submits job to the batch queue; (3) the user copies URL into browser window and waits for a node to be allocated; (4) when node is ready, user can access the notebooks; (5) when done, user shuts down notebook server.

As an example of using the Satellite Client as a system-wide resource,
the \textit{start-jupyter} was used for the 2020 SDSC Summer Institute. In this case, class accounts were configured with a default Miniconda environment, and we installed a global version of the script that was used by over 60 participants in the class to run Juptyer Notebeooks on CPU and GPU nodes. In addition, we are working with the Expanse User Portal team to add SJNPS to the portal as an interactive notebook service. 

For more details on using Satellite on the SDSC Expanse cluster, see the Notebooks 101 User Guide \cite{Notebooks101}, and for example notebooks that 
run on SDSC HPC systems, for both CPU and GPU devices, see the Notebook-Examples GitHub repository  \cite{github-ntbk-examp}.

\section{Conclusions and Future Plans}
The Satellite Proxy Service has been shown to be a reasonable solution to provide a way for our users to easily and securely launch single-user notebooks and to address to our security concerns. Satellite has been in production since Fall, 2020. Since then, it has been run on multiple SDSC HPC systems, spawned notebooks on both CPU and GPU nodes, and supports singularity containers. It has been used by individual users, and for training large groups. In general, users find the service easy to download and deploy, which is a major goal of the project.

Future plans include improving the functionality of the Satellite clients and service. On the client side, the \textit{start-jupyter} script encompasses most use cases, and is effective for a single-user installation. Based on user feedback, we have updated the pending (or loading) page interface to be more descriptive for users and to include basic job information. This is important for managing user expectations when the batch queue is busy it takes more than a few minutes to load the notebook. We are also working to improve two limitations of the script: make it easier to install client scripts as a system-wide resource, so Satellite clients can be run from any location; add to allow users to run notebooks from any location. 

Additionally, we have been developing a more advanced Satellite client called \textit{galyleo}, which expands on the capabilities and functions of the \textit{start-jupyter} script, and solves many usage scenario requirements. Enhanced features include: building the batch script in memory; passing the name and location of a desired conda environment or singularity container; greater control over memory, cpu and gpu settings; dynamically built batch scripts (which are saved in the users directories). One feature that has significant potential is the ability for a user to run \textit{galyleo} remotely (e.g. from a linux terminal on a local laptop), and generate a secure URL connection to a notebook running on an HPC system without needing to log directly onto the HPC system. We are using this feature with the Expanse User Portal team to use Satellite and the \textit{galyleo} script to spawn single-user notebooks from the portal. 

Future plans also include the addition of unit testing using Travis CI Test system (a service used to build and test software projects hosted on GitHub and Bitbucket 
\cite{TravisCI})
; exploring the use of Internal public-key-infrastructure for enabling TLS/HTTPS between Satellite and proxied applications; the development of usage metrics; and updating the Satellite service installation and deployment architecture.

\section{Acknowledgements}
The authors would like to thank the members of the SDSC Operations and User Services Groups, and Richard Wagner (UCSD IT Services) for their input and support, as well as early testers and attendees of the SDSC Summer Institute for providing testing and feedback. The work on this project was carried out with the support of the San Diego Supercomputer Center, the NSF Funded 
CC* Compute: Triton Stratus (\#1925558),  and NSF funded Expanse project (\#1928224) and the Extreme Science and Engineering Discovery Environment (NSF award \#ACI-1548562).

\bibliographystyle{unsrt}
\bibliography{srps-arxiv}

\end{document}